\documentclass[aps,pre,twocolumn, showpacs,amsmath,amssymb]{revtex4-1}
\usepackage[final]{graphicx}
\graphicspath{{./figs/}}
\usepackage{bm}
\usepackage{enumerate}
\usepackage{xcolor}

\usepackage{hyperref}
\hypersetup{
    pdfauthor={Navdeep Rana, Prasad Perlekar},
    pdftitle={Coarsening in the two-dimensional incompressible Toner-Tu equation: Signatures of turbulence}
}

\usepackage[capitalise]{cleveref}

\def \Rey {\mbox{Re}}
\def \Cn  {\mbox{Cn}}
\def \kL  {k_\mathcal{L}}
\def \klc {k_{\ell_c}}

\newcommand{\REM}[1]{{}}

\begin{document}
\title{Coarsening in the two-dimensional incompressible Toner-Tu equation: Signatures of turbulence}
\author{Navdeep Rana}
\author{Prasad Perlekar}
\affiliation{Tata Institute of Fundamental Research, Centre for Interdisciplinary Sciences, Hyderabad, India}

\begin{abstract}
    We investigate coarsening dynamics in the two-dimensional, incompressible Toner-Tu equation. We show that coarsening
    proceeds via vortex merger events, and the dynamics crucially depend on the Reynolds number $\Rey$. For low $\Rey$, the
    coarsening process has similarities to Ginzburg-Landau dynamics. On the other hand, for high $\Rey$, coarsening shows
    signatures of turbulence. In particular, we show the presence of an enstrophy cascade from the intervortex separation scale
    to the dissipation scale.
\end{abstract}

\maketitle

\section{\label{sec:intro} Introduction}

Active matter theories have made remarkable progress in understanding the dynamics of suspension of active polar particles (SPP)
such as fish schools, locust swarms, and bird flocks \cite{ram10,mar13,sri19r}. The particle based Vicsek model \cite{tam95} and
the hydrodynamic Toner-Tu (TT) equation \cite{ton98} provide the simplest setting to investigate the dynamics of SPP. Variants
of the TT equation have been used to model bacterial turbulence \cite{wensink2012,linkmann2019} and pattern formation in active
fluids \cite{sum04,gowrishankar2016,mar16,hus17,alert2020}. An important prediction of these theories is the presence of a
liquid-gas-like transition from a disordered gas phase to an orientationally ordered liquid phase \cite{ram10,cat15,cha20}. This
picture is dramatically altered if the density fluctuations are suppressed by imposing an incompressibility constraint. Toner
and colleagues \cite{che15,che16}, using dynamical renormalization group studies, showed that for the incompressible Toner-Tu
(ITT) equation the order-disorder transition becomes continuous. The near ordered state of the wet SPP on a substrate or under
confinement \cite{bricard2013, che16, maitra2020} belongs to the same universality class as the two-dimensional (2D) ITT
equation.

Investigating coarsening dynamics from a disordered state to an ordered state in systems showing phase transitions has been the
subject of intense investigation \cite{kib80,bra94,chu91,dam96,Puri09,per19}. In active matter coarsening has been studied
either in systems showing motility-induced phase separation \cite{cat15,tiri15} or for dry aligning dilute active matter (DADAM)
\cite{cha20,mish2010,das2018,kat20}. A key challenge in understanding coarsening in DADAM comes from the fact that the density
and the velocity field are strongly coupled to each other. Indeed, Ref.~\cite{mish2010} used both the density and the velocity
correlations to study coarsening in the TT equation. The authors observed that the coarsening length scale grew faster than
equilibrium systems with the vector order parameter and argued that the accelerated dynamics are because of the advective
nonlinearity in the TT equation. However, how nonlinearity alters energy transfer between different scales remains unanswered.

The incompressible limit, where the velocity field is the only dynamical variable, provides an ideal platform to investigate the
role of advection. Therefore, in this paper, we investigate coarsening dynamics using the ITT equation \cite{che16}:
\begin{equation}\label{eq:tonertu} \begin{aligned}
    \partial_{t}\bm{u}+\lambda\bm{u}\cdot\nabla\bm{u} = -\nabla P + \nu\nabla^2 \bm{u} + \bm{f},
\end{aligned} \end{equation}
where $\bm{u}(\bm{x},t)$ is the velocity field at position $\bm{x}$ and time $t$, $\lambda$ is the advection coefficient, $\nu$
is the viscosity, $\bm{f} \equiv \left(\alpha - \beta |\bm{u}|^2\right)\bm{u}$ is the active driving term with coefficients
$\alpha, \beta>0$, and the pressure $P({\bm x},t)$ enforces the incompressibility criterion $\nabla\cdot\bm{u}=0$. We do not
consider the random driving term in \eqref{eq:tonertu} because we are interested in coarsening under a sudden quench to zero
noise. For $\lambda=0$ and in the absence of the pressure term, \eqref{eq:tonertu} reduces to the Ginzburg-Landau (GL) equation.
On the other hand, \eqref{eq:tonertu} reduces to the Navier-Stokes (NS) equation on fixing $\alpha=0$, $\beta=0$, and
$\lambda=1$. Since most studies of dry active matter are done on a substrate, we investigate coarsening in two space dimensions.

By rescaling $\bm{x} \to \bm{x}/L$ , $t \to \alpha t$, $\bm{u} \to \bm{u}/U$ and $P \to P/\alpha U L$, we find that the Reynolds
number $\Rey=\lambda UL/\nu$ and the Cahn number $\Cn={\ell_c}/L$ completely characterize the flow (see
\cref{sec:nondimensional}). Here $U=\sqrt{\alpha/\beta}$ is the characteristic speed, and ${\ell_c} =\sqrt{\nu/\alpha}$ is the
length scale above which fluctuations in the disordered state $\bm{u}=0$ are linearly unstable.

We use a pseudospectral method \cite{per09,per10} to perform direct numerical simulation (DNS) of \eqref{eq:tonertu} in a
periodic square box of length $L$. The simulation domain is discretized with $N^2$ collocation points. We use a second-order
exponential time differencing (ETD2) scheme \cite{cox02} for time marching. Unless stated otherwise, we set $L=2\pi$ and
$N=2048$. We initialize our simulations with a disordered configuration, randomly oriented velocity vectors drawn from a
Gaussian distribution with zero mean and standard deviation $\sigma= U/3$, and monitor the coarsening dynamics. Our main
findings are as follows:
\begin{enumerate}[(i)]
    \item Coarsening proceeds via vortex mergers.
    \item For low $\Rey$, advective nonlinearities can be ignored, and the dynamics resembles coarsening in the GL equation.
    \item For high $\Rey$, we find signatures of 2D turbulence, and the coarsening accelerates with increasing $\Rey$. We also
        provide evidence of a forward enstrophy cascade which is a hallmark of 2D turbulence.
\end{enumerate}

\begin{figure*}
    \includegraphics[width=\linewidth]{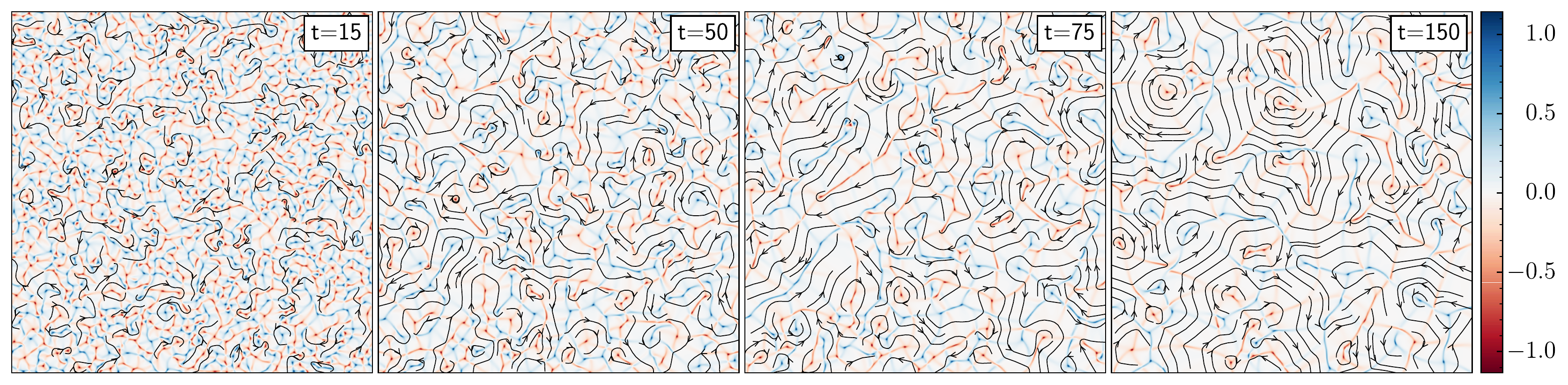}
    \put(-531,110){\Large (a)}

    \includegraphics[width=\linewidth]{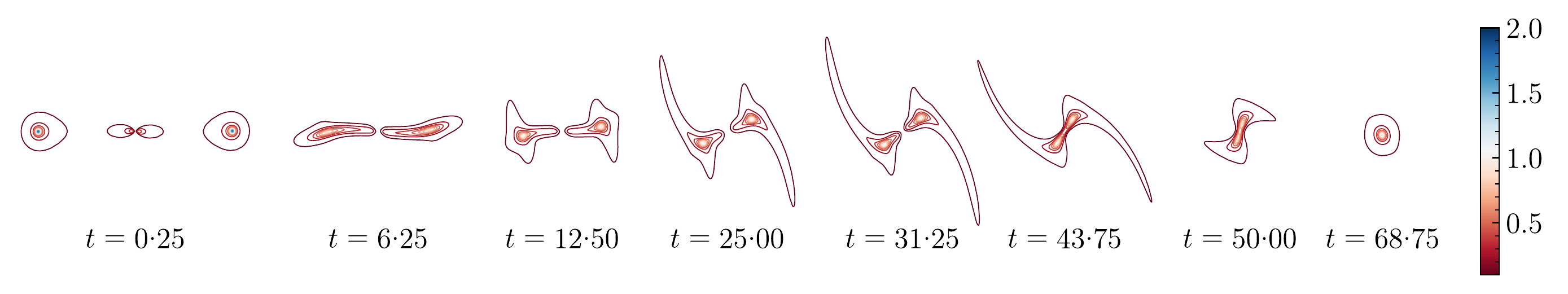}
    \put(-533,075){\Large (b)}

    \includegraphics[width=\linewidth]{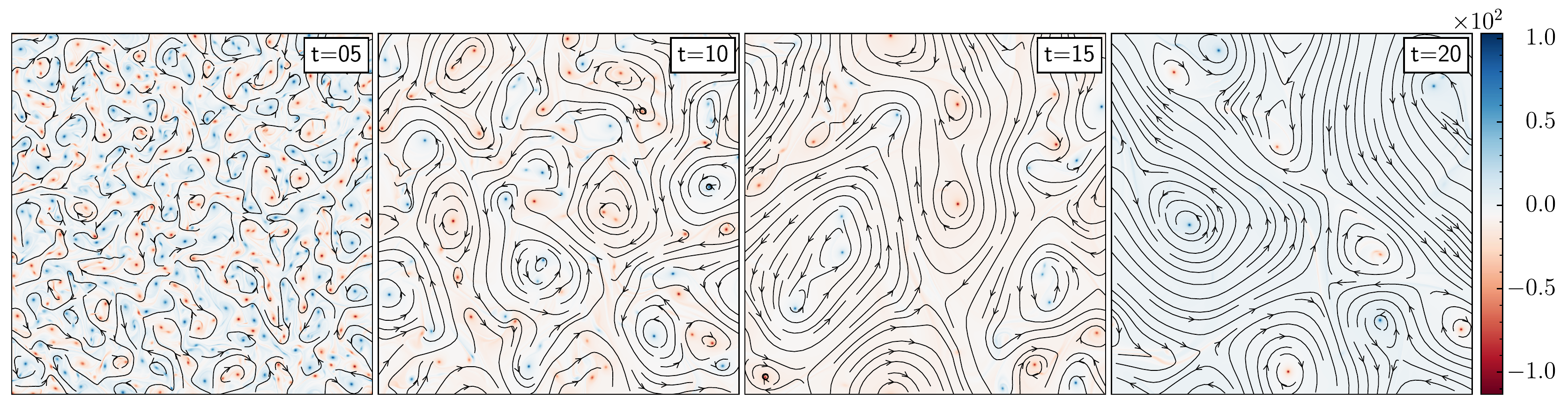}
    \put(-531,110){\Large (c)}

    \includegraphics[width=\linewidth]{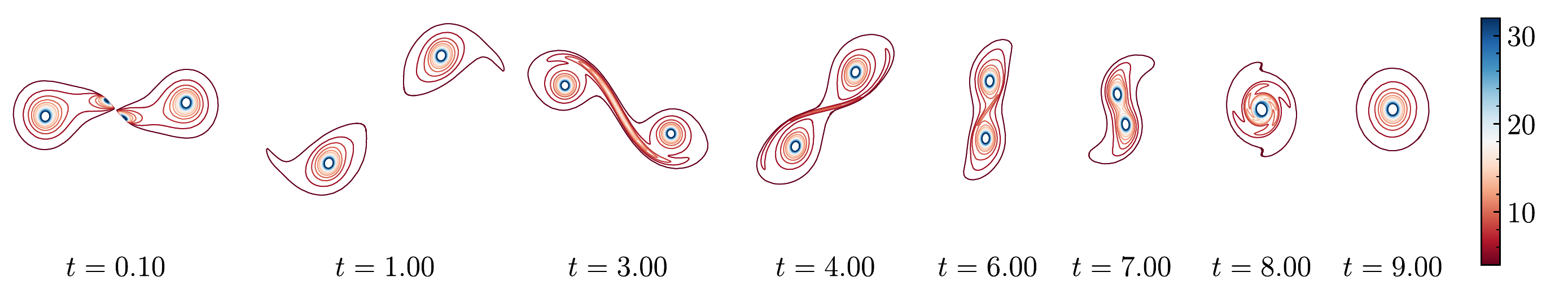}
    \put(-533,078){\Large (d)}

    \caption{\label{fig:2dsnapshot}
        Pseudocolor plots of the vorticity field $\omega=\hat{z} \cdot \nabla \times \bm{u}$ superimposed on the velocity
        streamlines at different times for (a) $\Rey=2\pi\times10^2$ and (c) $\Rey=2\pi \times 10^{4}$ in the coarsening regime.
        Contour plots of the vorticity field $\omega$ showing the merger of two isolated corotating vortices
        (vortex-saddle-vortex configuration) at (b) $\Rey=2\pi\times10^2$ and (d) $\Rey=2\pi \times 10^{4}$.
    }
\end{figure*}

In the following sections we discuss our results on the coarsening dynamics and then present conclusions in \cref{sec:concl}.

\section{Results}

In the following, we quantity how the vortex dynamics controls coarsening. The pseudocolor plot of the vorticity field in
\cref{fig:2dsnapshot}(a) and (c) shows different stages of coarsening at low $\Rey=2\pi \times 10^2$ and high
$\Rey=2\pi\times10^4$. During coarsening, vortices merge and the inter-vortex spacing continues increasing. For low $\Rey=2\pi
\times 10^{2}$ [see \cref{fig:2dsnapshot}(a)], the dynamics in the coarsening regime resembles defect dynamics in the
Ginzburg-Landau equation \cite{bra94,Onu02, Puri09}. On the other hand, for high $\Rey=2\pi \times 10^{4}$, vorticity snapshots
resemble 2D turbulence. In particular, similar to vortex merger events in 2D \cite{meunier2005,dizes2002,lew16,swaminathan2016},
it is easy to identify a pair of corotating vortices undergoing a merger and the surrounding filamentary structure. Earlier
studies\cite{niel96,kev97,swaminathan2016} on the vortex merger in two-dimensional Navier-Stokes equations showed that the
filamentary structures formed during the merger process lead to an enstrophy cascade. Because the ITT equation structure is
similar to NS equations, we expect that the vortex merger at high $\Rey$ will also lead to an enstrophy cascade. 

To further investigate the vortex merger, we perform DNS of the isolated vortex-saddle-vortex configuration at various Reynolds
numbers. For these simulations we use $N=4096$ collocation points. Furthermore, to minimize the effect of periodic boundaries,
we set $\alpha=-10$ for $r>0.9 L/2$ and keep $\alpha=1$ otherwise, where $r\equiv \sqrt{(x - L/2)^2 + (y-L/2)^2}$. This ensures
that the velocity decays to zero for $r\geq0.9L/2$. Note that a vortex in the 2D ITT equation is a point defect with unit
topological charge and core radius ${\ell_c}$ (see \cref{sec:vortex}).

We observe that during the evolution of a vortex-saddle-vortex configuration [see \autoref{fig:2dsnapshot}(b) and (d)]: (i)
similar to defect dynamics in the GL equation \cite{yurke93,chai98,Onu02}, each vortex gets attracted to the saddle due to the
opposite topological charge, (ii) the two vortices rotate around each other, similar to convective merging in NS
\cite{swaminathan2016,lew16}, and (iii) the flexure of the vortex trajectory depends on $\Rey$ (see \cref{sec:merger}). Thus, a
vortex merger event in the two-dimensional ITT equation has ingredients from both the NS and GL equations. In \cref{sec:merger},
we provide a more detailed investigation of the vortex merger with varying $\Rey$.

To quantify coarsening dynamics, we conduct a series of high-resolution DNSs ($N=2048$) of the ITT equation by varying $\Rey$
while keeping $\Cn=1/(100 L)$ fixed. For ensemble averaging, we evolve $48$ independent realizations at every $\Rey$. We monitor
the evolution of the energy spectrum
$E_k(t) \equiv \frac{1}{2}\sum_{k-1/2\leq p < k+1/2} \langle |\hat{\bm{u}}_{\bm{p}}(t)|^2
\rangle$, and the energy dissipation rate (or equivalently the excess free
energy) $\epsilon(t) \equiv \left< 2 \nu \sum_k k^2 E_k(t) \right>$. Here
$\hat{\bm{u}}_{\bm{k}}(t)\equiv \sum_{\bm{x}} \bm{u}(\bm{x},t) \exp(-i \bm{k}\cdot\bm{x})$, $i=\sqrt{-1}$, and
the angular brackets indicate the ensemble average \footnote{ The energy spectrum $E_k$ and the structure factor
$S_k$ are related to each other as $E_k=k^{d-1} S_k$. }.

\subsection{Energy Dissipation Rate}

\begin{figure}
    \includegraphics[width=\linewidth]{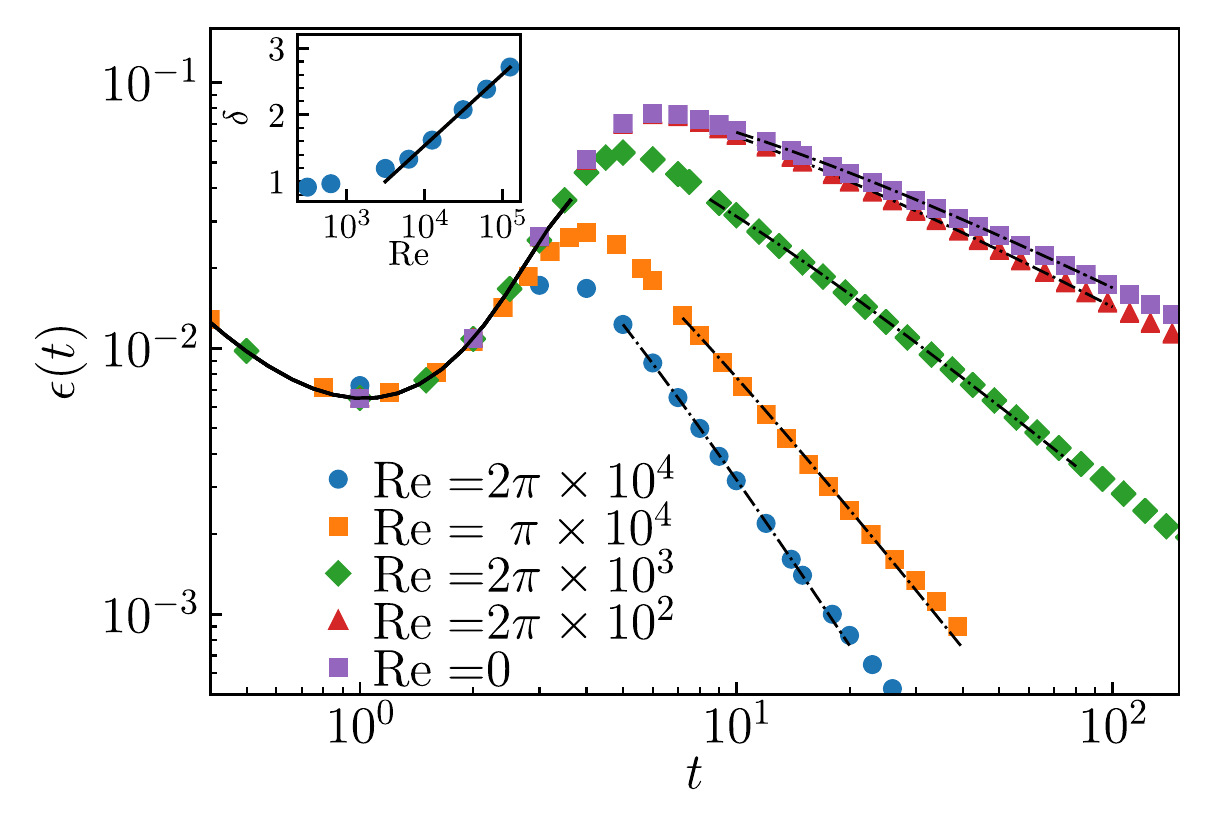}
    \caption{ \label{fig:dissipation}
        Plot of the energy dissipation rate $\epsilon(t)$ vs time at various Reynolds numbers. The early time evolution of
        $\epsilon(t)$ is well approximated by \eqref{eq:norm} (solid black line). At late times, $\epsilon(t)$ decays as
        $\epsilon(t) \sim t^{-\delta}\ln(t)$ (black solid lines), with $\delta$ obtained using a least-squares fit. Inset: Plot
        of $\Rey$ vs $\delta$ and the fit $\delta \sim -2.71 + 0.46 \ln(\Rey)$ for $\Rey >> 1$. For $\Rey \rightarrow 0$,
        consistent with Ginzburg-Landau scaling, we obtain $\delta \to 1$. 
    }
\end{figure}

The time evolution of the energy dissipation rate $\epsilon(t)$ is shown in \cref{fig:dissipation}. For the initial disordered
configuration, because the statistics of velocity separation is Gaussian, we approximate the fourth-order correlations in terms
of the product of second-order correlations to get the following equation for the early time evolution of the energy
spectrum~\cite{bra15}:
\begin{equation}\begin{aligned}
    \partial_t E_k(t) \approx [2\alpha -8\beta E(t)]E_k(t) -2\nu k^2 E_k(t),
    \label{eq:norm}
\end{aligned}\end{equation}
where $E(t)=\sum_{k} E_k(t)$. In \cref{fig:dissipation} we show that the early-time evolution of the energy dissipation rate
$\epsilon(t)$ obtained from \eqref{eq:norm} is in good agreement with the DNS.

For late times, coarsening proceeds via vortex (defect) mergers. For GL equations in two dimensions, Refs.~\cite{yurke93,qian03}
show that $\epsilon(t) \propto t^{-1}\ln(t)$. In our simulations, we find that $\epsilon(t) \propto t^{-\delta}\ln(t)$, where
$\delta$ is now $\Rey$ dependent. For low $\Rey$, where the effect of the advective nonlinearity can be ignored, we recover GL
scaling ($\delta \rightarrow 1$ as $\Rey \rightarrow 0$). For high $\Rey$, coarsening dynamics is accelerated with $\delta \sim
-2.71 + 0.46\ln(\Rey)$ [see \cref{fig:dissipation}, inset]. 

\begin{figure}
    \centering
    \includegraphics[width=0.9\linewidth]{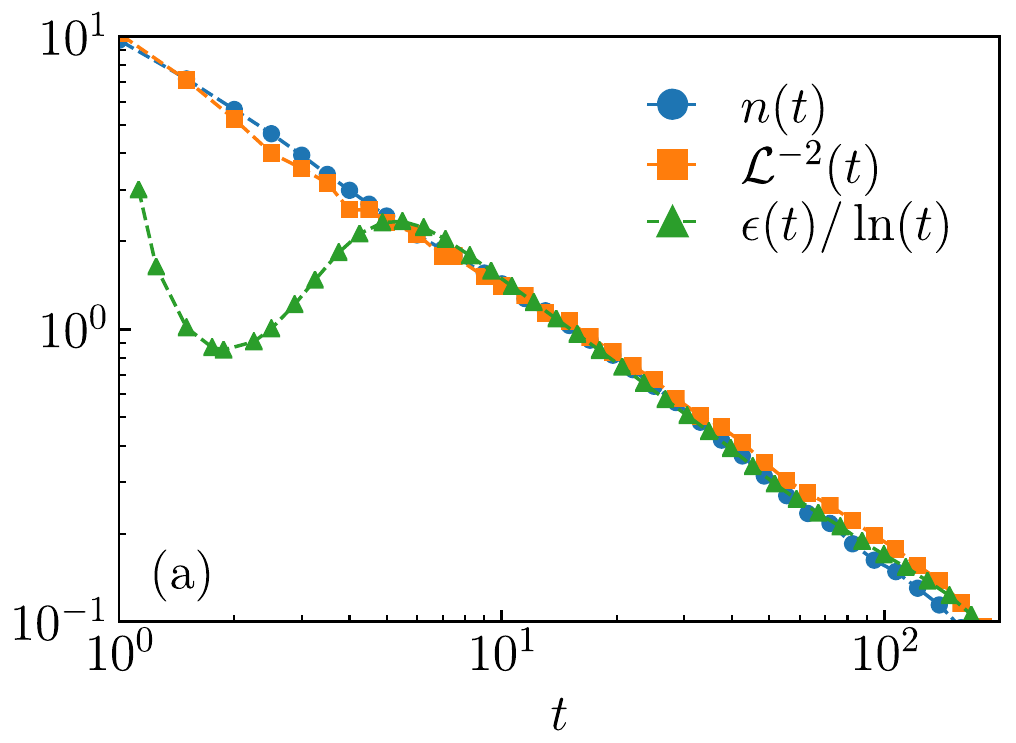}
    \includegraphics[width=0.9\linewidth]{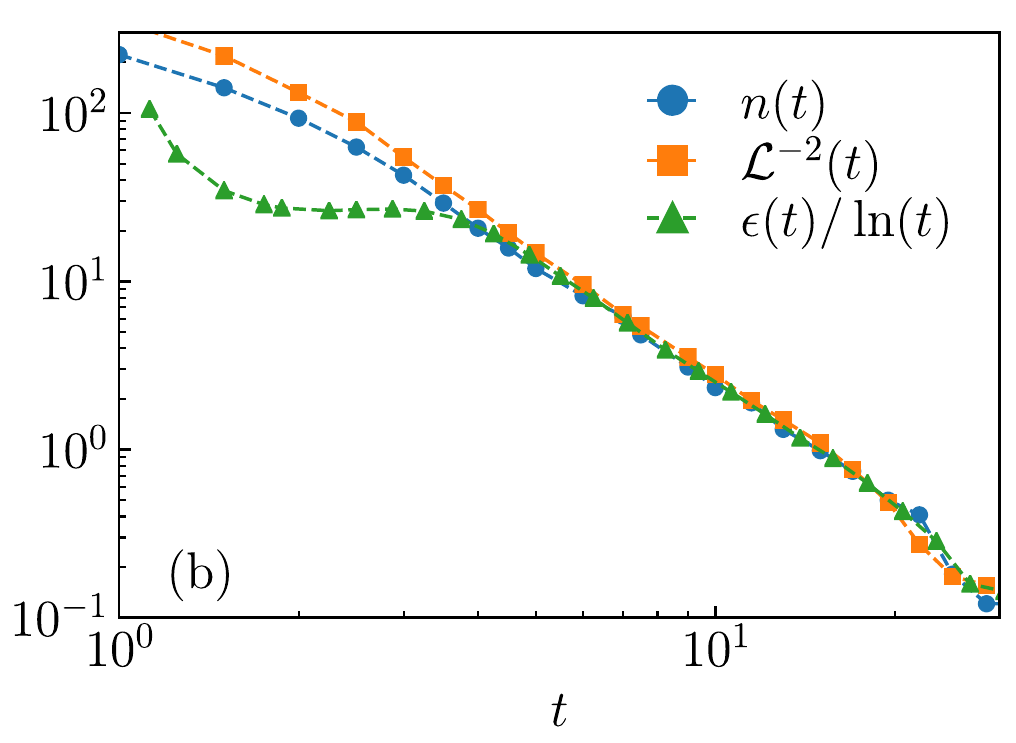}
    \caption{\label{fig:number}
        Plots comparing the time evolution of $n(t)$, $\mathcal{L}(t)$, and $\epsilon(t)$ for (a) $\Rey = 2\pi\times 10^2$, and
        (b) $\Rey = 2\pi\times 10^4$. The curves are vertically shifted to highlight identical scaling behavior [$n(t) \propto
        \mathcal{L}^{-2}(t) \propto \epsilon(t)\ln(t) \propto t^{-\delta}$] in the coarsening regime.
    }
\end{figure}

\subsubsection{\label{sec:dissdel} Energy dissipation rate and the coarsening length scale}

We now discuss the relationship between the energy dissipation rate, the defect number density, and the coarsening length scale.
The coarsening length scale \cite{Onu02,Puri09,per14,per17,per19} 
\begin{equation}
    \mathcal{L}(t) \equiv 2 \pi \frac{\sum_k E_k(t)}{\sum_k k E_k(t)}
\end{equation}
has been used to monitor inter-defect separation during the dynamics. 

We identify defects from the local minima of the $|{\bm u}|$ field in our DNS of the ITT equation and define the defect number
density as $n(t)\equiv {\cal N}_d(t)/L^2$, where ${\cal N}_d$ denotes the number of defects at time $t$ \footnote{We use
scikit-image library \cite{scikit-image} to identify local minima}. In \cref{fig:number}, we show that in the coarsening regime
$n(t) \propto \mathcal{L}^{-2}(t) \propto \epsilon(t)/\ln(t)$ for low $\Rey=2\times 10^{2}$ as well as high $\Rey=2 \times
10^{4}$. As discussed above, the energy dissipation rate decays as $\epsilon(t)\sim t^{-\delta}\ln(t)$ in the coarsening regime.
Similar to GL dynamics, we find that $n(t) \propto \mathcal{L}^{-2}(t)$ even for the ITT equation. However, both $n(t)$ and
$\mathcal{L}^{-2}(t)$ show a power-law decay ($n \propto {\mathcal L}^{-2} \sim t^{-\delta}$) without any logarithmic
correction.

\begin{figure}
    \centering
    \includegraphics[width=0.95\linewidth]{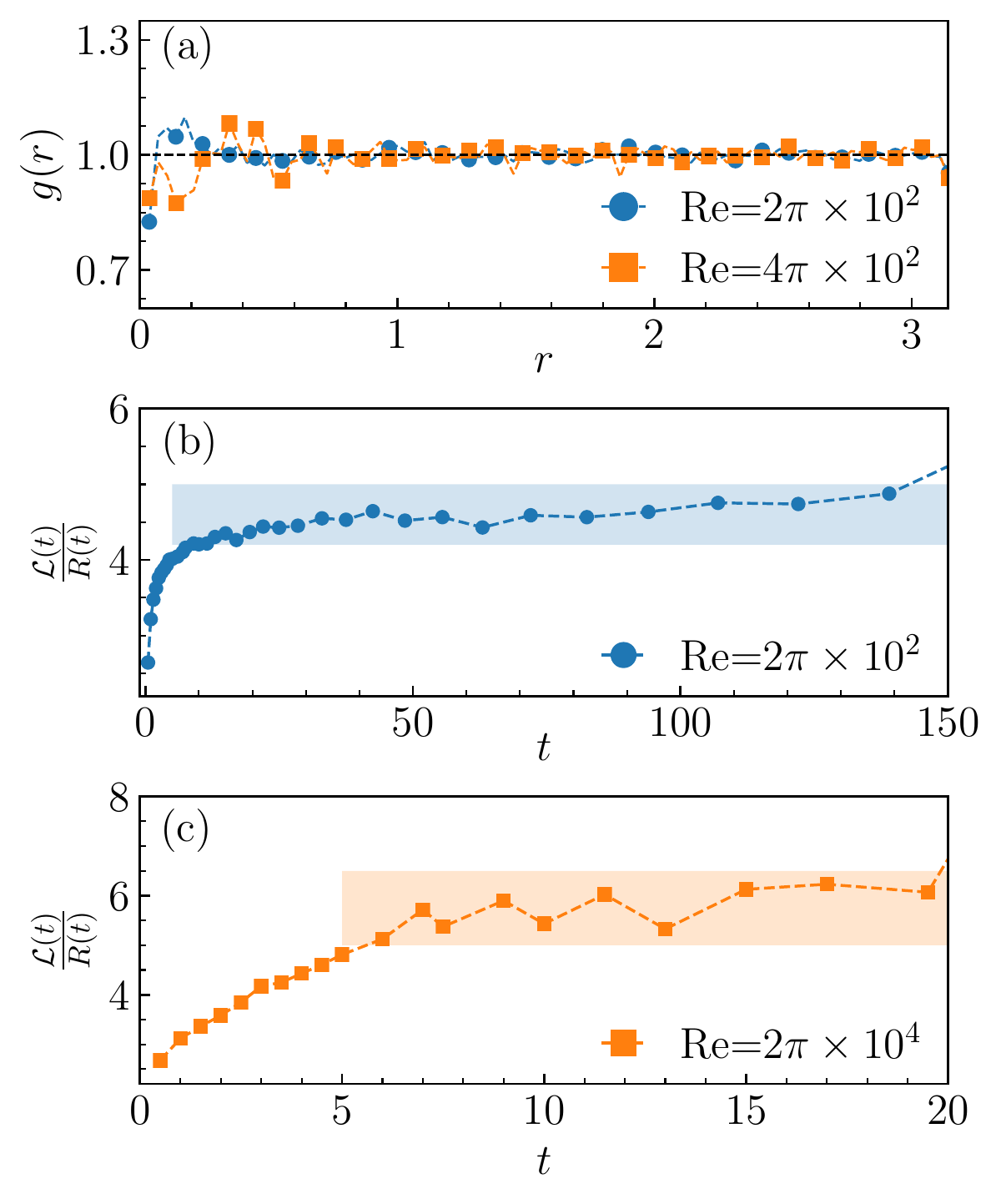}
    \caption{\label{fig:geometric}
        (a) Plot of the radial distribution function $g(r)$ for $\Rey = 2\pi\times 10^2$ at time $t=40$ and $\Rey = 2\pi\times
        10^4$ at time $t=10$ in the coarsening regime. The dashed black line indicates theoretical prediction $g(r)=1$ for
        uniformly distributed points. Plots showing $\mathcal{L}(t)/R(t)$ for (b) $\Rey = 2\pi\times 10^2$ and (c) $\Rey =
        2\pi\times 10^4$. $\mathcal{L}(t)/R(t)$ is fairly constant in the coarsening regime (shaded region).
    }
\end{figure}

A purely geometrical argument can be constructed to explain the observed relation between $n(t)$ and $\mathcal{L}(t)$. As we
start our simulations from a disordered configuration, defects are expected to be uniformly distributed over the entire
simulation domain. In \cref{fig:geometric}(a), we plot the radial distribution function \cite{allen2017}
\begin{equation}
        g(r) \equiv \frac{1}{2\pi r dr n(t)}\sum_{i\neq j}\delta(r-r_{ij}).
\end{equation} 
Here $r_{ij} = |\bm{r}_{i} - \bm{r}_{j}|$, $\bm{r}_{i}$ are the defect coordinates and $ dr$ is the bin width used to calculate
$g(r)$. Consistent with our assumption above, we find $g(r)=1$, indicating defects are uniformly distributed in the coarsening
regime. Then following Refs.~\cite{Cha43,her1909} we get $R(t) = 1/2\sqrt{n(t)}$, where $R(t)$ is the average nearest-neighbor
distance at time $t$. Consistent with the dynamic scaling hypothesis \cite{bra94}, in \cref{fig:geometric}(b) and (c) we show that
${\mathcal L}(t) \propto R(t)$ in the coarsening regime. Using this, we get ${\mathcal L}(t) \propto 1/\sqrt{n(t)}$ independent
of $\Rey$. 

For systems with topological defects, the energy dissipation rate (or the excess free energy) is proportional to the defect
number density $n(t)$ \cite{bra94,lubensky1995,chai98,yurke93,qian03}. Thus, consistent with \cref{fig:number}, we get ${\cal
L}(t) \propto 1/\sqrt{\epsilon (t)}$ (apart from the logarithmic factor).

\begin{figure}
    \includegraphics[width=\linewidth]{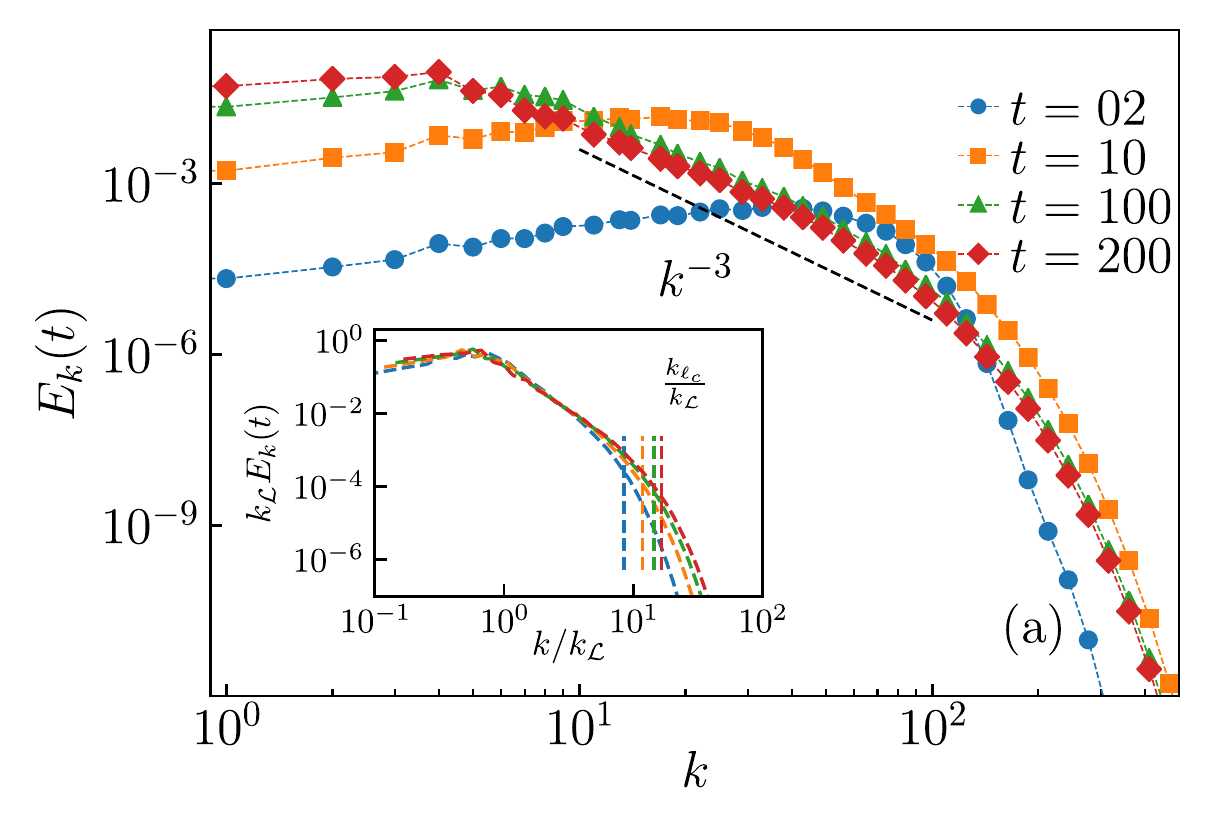}
    \includegraphics[width=\linewidth]{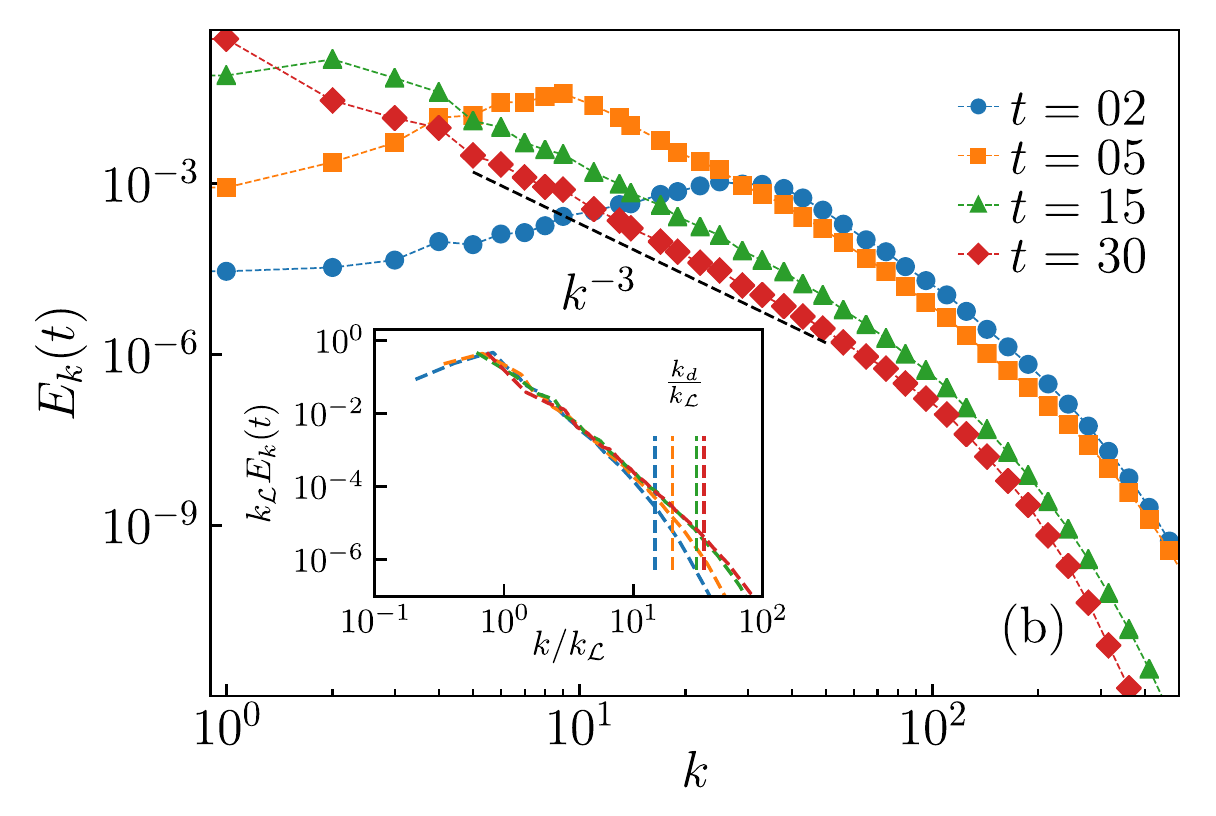}
    \caption{ \label{fig:spectrum}
        Time evolution of the energy spectra for (a) $\Rey=2\pi\times10^2$, and (b) $\Rey=2\pi\times10^4$. Inset: The scaled
        energy spectrum $\kL E_k(t)$ versus $k/\kL$ shows an excellent collapse between wave numbers $\kL$ and $\klc (k_d)$ for
        $\Rey=2\pi \times 10^{2}$ ($\Rey=2\pi \times 10^{4}$), confirming the dynamical scaling hypothesis. The wave numbers
        $\klc$ and $k_d$ at different times are marked by vertical dashed lines (same colors as the spectra).
    }
\end{figure}

\subsection{Energy spectrum}

The plots in \cref{fig:spectrum}(a) and (b) show the energy spectrum $E_k(t)$ versus $k$ at different times for low $\Rey= 2\pi
\times 10^2$ and high $\Rey=2\pi\times10^4$. In both cases, the energy spectrum in the coarsening regime shows a power-law scaling $E_k(t)
\propto k^{-3}$. We find that consistent with the dynamic scaling hypothesis \cite{bra94}, the scaled spectrum collapses between
wave numbers $\kL \equiv 1/\mathcal{L}$ and $\klc \equiv \ell_c^{-1}$ for low $\Rey$. At high $\Rey$ the collapse is between
$\kL$ and the dissipation wave number $k_d$ [see insets in \cref{fig:spectrum}(a) and (b)].

The observed $k^{-3}$ scaling for the energy spectrum can appear because of (i) the modulation of the velocity field around the
topological defects (Porod's tail) \cite{Onu02} and (ii) the enstrophy cascade, similar to two-dimensional turbulence, due to
the advective nonlinearity in \eqref{eq:tonertu}.

\subsection{Enstrophy Budget}

To investigate the dominant balances between different scales, we use the scale-by-scale enstrophy budget equation
\begin{equation} \label{eq:ensb}
    \partial_t \Omega_k(t) + T_k(t) = -2 \nu k^2 \Omega_k(t) + {\mathcal F}_k(t),
\end{equation}
where $\Omega_k \equiv k^2 E_k$ is the enstrophy, $\mathcal{F}_{k}(t) \equiv k^2(\hat{\bm{u}}_{-k}\cdot\hat{\bm{f}}_k
+\hat{\bm{u}}_{k}\cdot\hat{\bm{f}}_{-k})$ is the net enstrophy injected because of active driving, $T_k \equiv d Z_k(t)/dt$ is
the enstrophy transfer function, and $Z_{k}
\equiv \sum_{|\bm{p}|\le|{\bm k}|}^{N/2} \hat{\omega}_{\bm{p}} \cdot \widehat{(\bm{u}\cdot \nabla
\omega)}_{-\bm{p}}$ is the enstrophy flux.

\begin{figure}
    \includegraphics[width=\linewidth]{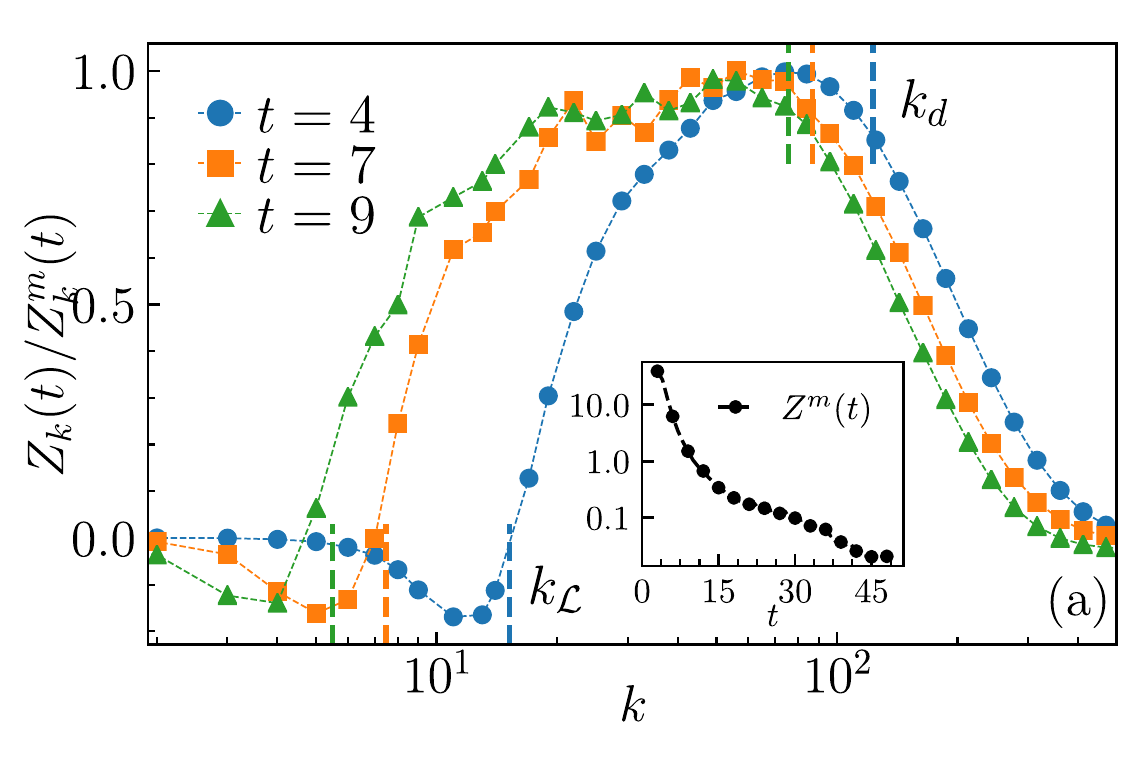}
    \includegraphics[width=\linewidth]{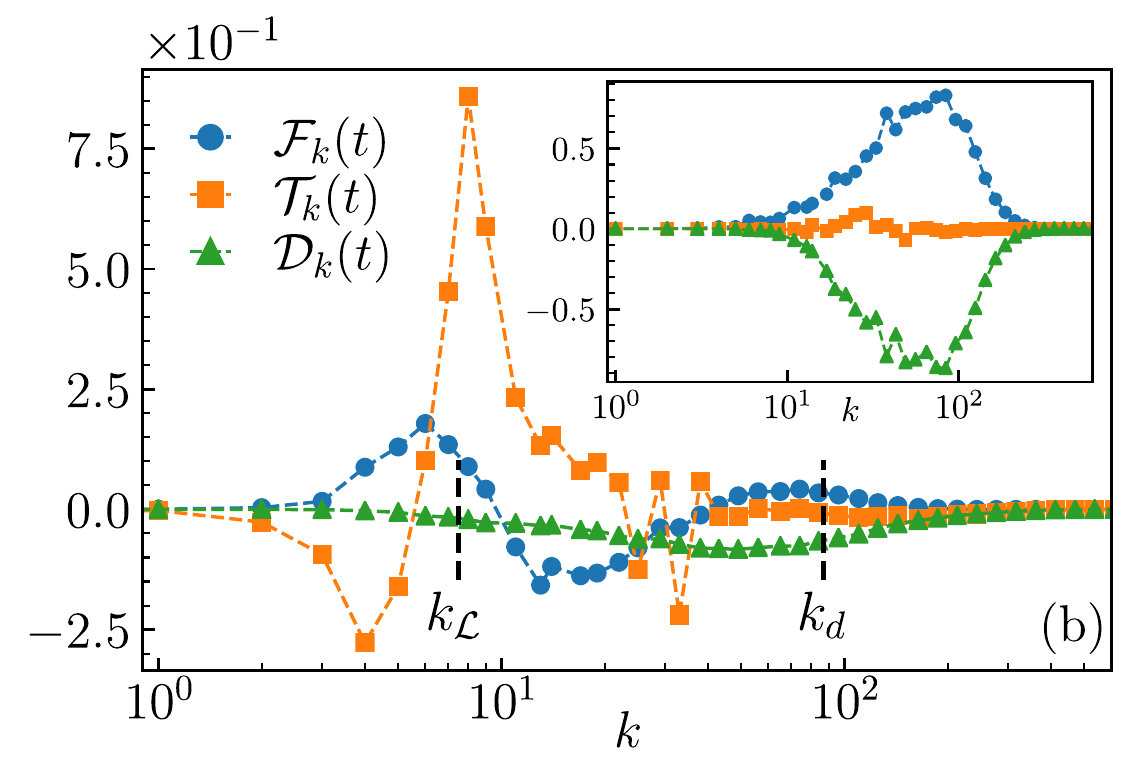}
    \caption{\label{fig:enstrophy}
        (a) Plot of the enstrophy flux $Z_k(t)/Z^m(t)$ versus $k$ at $\Rey=2\pi\times10^4$ for different times in the coarsening
        regime. Wave numbers $\kL$ and ${k_d}$ are marked with vertical dashed lines (same colors as the main plot). Inset: Time
        evolution of $Z^m(t)$. (b) Enstrophy budget: Plot of the transfer function ${\mathcal T}_k\equiv dZ_k/dk$, enstrophy
        injection due to the active driving ${\mathcal F}_k$, and the enstrophy dissipation $\mathcal{D}_k = -2 \nu k^2
        \Omega_k$ for $\Rey = 2\pi\times 10^4$ and at time $t=7$ in the coarsening regime. Inset: Plot of different terms in
        the enstrophy budget for low $\Rey = 2\pi \times 10^2$ and at time $t=25$ in the coarsening regime.
    }
\end{figure}

The classical theory of 2D turbulence \cite{kraic67,lei68,bat69,per09b,boffetta2012,pan17} assumes the presence of an inertial
range with constant enstrophy flux at scales smaller than the forcing scale and larger than the dissipation scale. Indeed, for
high $\Rey=2\pi \times 10^{4}$, in \cref{fig:enstrophy}(a) we confirm the presence of a positive enstrophy flux $Z_{k}$ between
wave number $\kL\equiv 1/{\mathcal{L}}$, corresponding to the intervortex, separation and the dissipation wave number $k_d \equiv
{\left({8\nu^3}/{Z^m}\right)}^{-1/6}$ for $2 \leq t <30$ in the coarsening regime. As the coarsening proceeds, the region of
positive flux becomes broader, and $\kL$ shifts to smaller wave numbers, but the maximum value of the flux $Z^m(t)$ decreases
[\cref{fig:enstrophy}(a), inset]. In \cref{fig:enstrophy}(b) we plot different terms in the enstrophy budget equation
\eqref{eq:ensb}. We find that the active driving primarily injects enstrophy (${\mathcal F}_k>0$) around wave number $\kL$ but,
unlike classical turbulence, it is not zero in the region of constant enstrophy flux ($\kL < k < k_d$). Viscous dissipation is
active only at small scales $k\geq k_d$. At late times $t>30$, the enstrophy flux is negligible [\cref{fig:enstrophy}(a,inset)].

For low $\Rey$, the enstrophy transfer $T_k$ is negligible, and the enstrophy dissipation $\mathcal{D}_k(t)$ balances the
injection because of the active driving ${\mathcal F}_k(t)$ [see \cref{fig:enstrophy}(b), inset]. Therefore, the $k^{-3}$ scaling
in the energy spectrum [\cref{fig:spectrum}(a)] is due to Porod's tail.

\begin{figure}[!t]
    \centering \includegraphics[width=\linewidth]{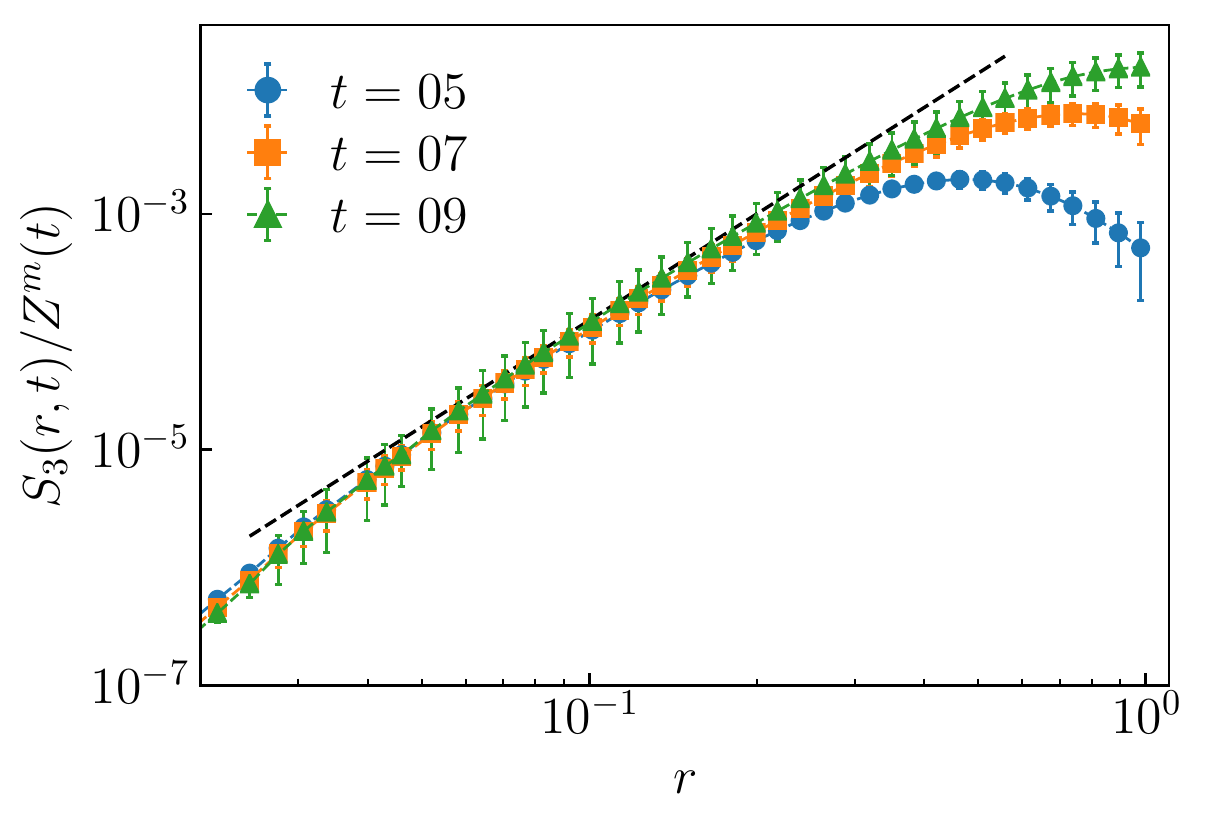}
    \caption{\label{fig:structure}
        Plot of the third-order velocity structure function $S_{3}(r,t)$ scaled by the maxima of enstrophy flux $Z^m(r)$ for $t=5-9$
        in the coarsening regime. The dashed black line shows the theoretical prediction $S_{3}(r)/Z^{m}(t) = \frac{1}{8}r^3$
        for comparison.
    }
\end{figure}

\subsection{Third-order Velocity Structure Function}

The real-space indicator of the enstrophy flux in 2D turbulence is the following exact relation for the third-order velocity
structure function:
\begin{equation} \label{eq:stf}
    S_{3}(r,t) = \frac{1}{8}Z_{k\sim1/r} r^3.
\end{equation}
Here $S_{3}(r,t) \equiv \langle \left[\delta_r u \right]^3\rangle$, $\delta_r u \equiv \left[\bm{u}(\bm{x}+\bm{r},t) -
\bm{u}(\bm{x},t)\right].\hat{\bm{r}}$, and the angular brackets indicate spatial and ensemble averaging \cite{cer17,lin96}. In
the statistically steady turbulence, the enstrophy flux $Z_k$ is constant in the inertial range and is equal to the enstrophy
dissipation rate. During coarsening in ITT, we observe a nearly uniform flux $Z_k$ for $\kL \leq k \leq k_{d}$, albeit with
decreasing magnitude [see \cref{fig:enstrophy}(a)]. Therefore, for ITT we choose $Z_{k\sim 1/r} = Z^{m}(t)$ in \eqref{eq:stf}.
In \cref{fig:structure}, we show the compensated plot of $S_3(r,t)$ in the coarsening regime and find the inertial range scaling
to be consistent with the exact result \eqref{eq:stf}.

\subsection{Effect of noise on the coarsening dynamics}
\begin{figure}
    \centering
    \includegraphics[width=\linewidth]{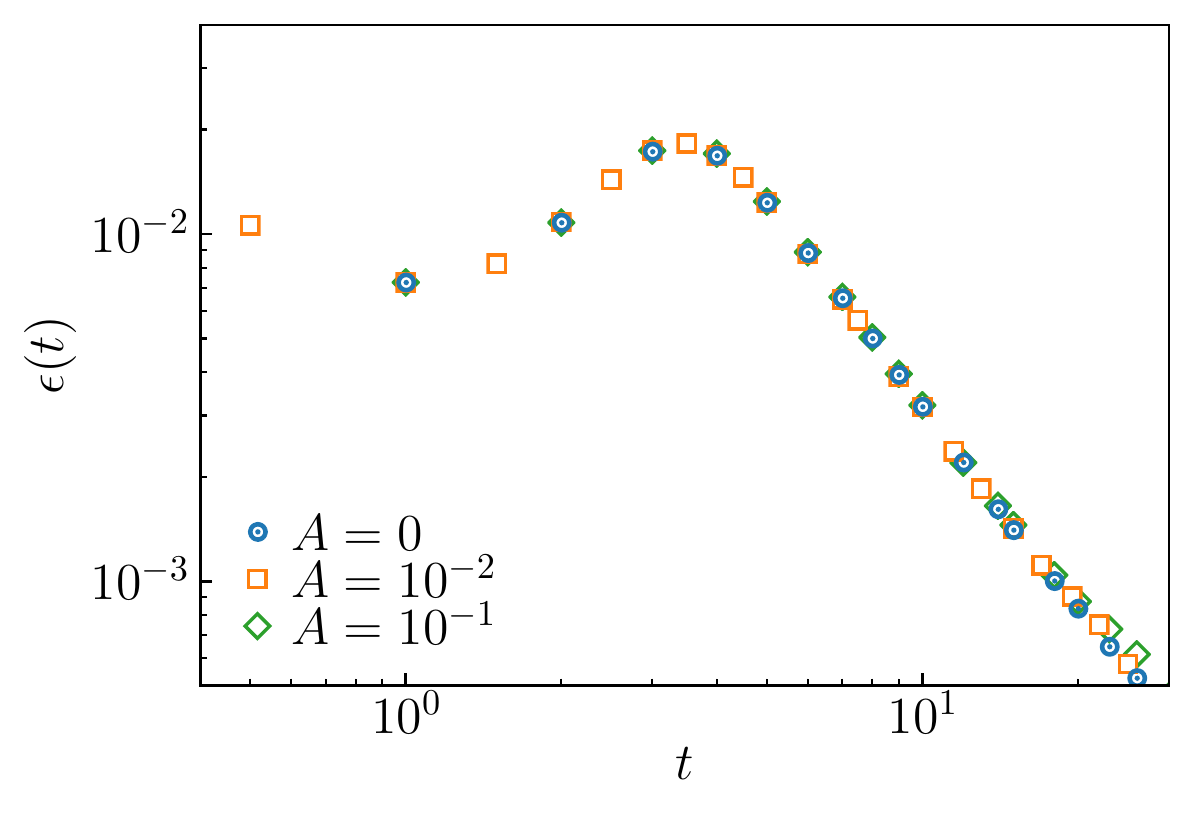}
    \caption{\label{fig:noise}
        Plot comparing the evolution of the energy dissipation rate at different noise strengths for $\Rey = 2\pi\times 10^{4}$.
        For ensemble averaging, we evolve 16 independent realizations at $A=10^{-1}$ and $A = 10^{-2}$.
    }
\end{figure}

To investigate the effect of noise on the coarsening dynamics, we add a Gaussian noise ${{\bm \eta}({\bm x},t)}$ to the ITT
equation \cite{che16},
\begin{equation}\label{eq:tonertunoise} \begin{aligned}
    \partial_{t}\bm{u}+\lambda\bm{u}\cdot\nabla\bm{u} = -\nabla P + \nu\nabla^2 \bm{u} + \bm{f} + \bm{\eta}, 
\end{aligned} \end{equation}
where $\left<{{\bm \eta}(\bm{x},t)}\right> = 0$ and $\left<\eta_{i}(\bm{x},t)\eta_{j}(\bm{x}',t')\right> =
A\delta_{ij}\bm{\delta}(\bm{x}-\bm{x}')\delta(t-t')$, where $A$ controls the noise strength. In \cref{fig:noise}, we show that
the evolution of the energy dissipation rate $\epsilon(t)$ for $\Rey=2\pi\times10^{4}$, averaged over $16$ independent noise
realizations, remains unchanged for different values of $A=0,0.1,$ and $0.01$. Clearly, the presence of noise in the ITT
equation does not alter the coarsening dynamics.

\section{Coarsening in ITT versus bacterial turbulence}

Bacterial turbulence (BT) refers to the chaotic spatiotemporal flows generated by dense suspensions of motile bacteria
\cite{dom04,wensink2012}. The dynamics of a turbulent bacterial suspension is modeled by the ITT equation, albeit with the
viscous dissipation in ITT replaced with a Swift-Hohenberg-type fourth-order term to mimic energy injection due to bacterial
swimming \cite{dun13,wensink2012,bra15,linkmann2019,linkmann2020}, 
\begin{equation} \label{eq:bt} 
    \partial_t {\bm u} + \lambda {\bm u} \cdot \nabla {\bm u} = -\nabla P -\nu \nabla^2 {\bm u} + \Gamma \nabla^4 {\bm u} + {\bm f}, 
\end{equation} 
where $\nu>0$ and the parameter $\Gamma>0$.

In contrast to BT \eqref{eq:bt} , the ITT is a model of flocking dynamics. Indeed the homogeneous, ordered state is a stable
solution of the ITT \eqref{eq:tonertu} but not of BT \eqref{eq:bt}. Furthermore, \eqref{eq:bt} and its variants show an inverse
energy transfer from small scales to large scales, whereas during coarsening in ITT we observe a forward enstrophy cascade from
the coarsening length scale ${\mathcal L}$ to small scales. 

\section{Conclusion \label{sec:concl}}

In conclusion, we have investigated coarsening dynamics in ITT equations. We find that at low Reynolds number the dynamics is
similar to coarsening in the Ginzburg-Landau equation, whereas for high Reynolds numbers coarsening shows signatures of 2D
turbulence. Specifically, for high Reynolds numbers, we showed the presence of an enstrophy cascade which accelerates the
coarsening dynamics and verified the exact relation for the structure function. Our results would also be experimentally
relevant to a dense suspension of active polar particles that undergo a flocking transition, such as suspensions of active polar
rods \cite{kud08,kum14}. 

\begin{acknowledgments}
    We thank S. Ramaswamy and H. Chat\'{e} for discussions and are grateful for support from intramural funds at TIFR Hyderabad
    from the Department of Atomic Energy (DAE), India, and DST (India) Project No. ECR/2018/001135. 
\end{acknowledgments}

\appendix

\section{\label{sec:nondimensional} Dimensionless ITT equation}
Consider the incompressible Toner-Tu (ITT) equation
\begin{equation}\begin{aligned}
    \nonumber
    \partial_t\bm{u}+\lambda\bm{u}\cdot\nabla\bm{u} = -\nabla P + \nu\nabla^2 \bm{u} + \left(\alpha - \beta |\bm{u}|^2\right)\bm{u} .
\end{aligned}\end{equation}
By rescaling the space ${\bm x'} \to {\bm x}/L$, the time $t' \to \alpha t$, the pressure $P' \rightarrow {P}/{\alpha L U}$, and
the velocity field $\bm{u'} \to \bm{u}/U$, the ITT equation becomes
\begin{equation}\begin{aligned}
    \nonumber
    \alpha U\partial_{t'}{\bm{u'}}+\frac{\lambda U^2}{L}\bm{u'}\cdot\nabla'\bm{u'} =
    - \alpha U \nabla' P' + \frac{\nu U}{L^2}\nabla'^2 \bm{u'} \\
    + \left(\alpha - \beta U^2 |\bm{u'}|^2\right)U\bm{u'},
\end{aligned}\end{equation}
where $U^2=\alpha/\beta$. Ignoring the primed index for convenience, we arrive at the dimensionless form of the ITT equation:
\begin{equation}\label{eq:tonertuReCn}\begin{aligned}
    \nonumber
    \partial_t\bm{u}+\Rey \Cn^2\bm{u}\cdot\nabla\bm{u} = -\nabla P + \Cn^2 \nabla^2 \bm{u}
    + \left(1 - |\bm{u}|^2\right)\bm{u}.
\end{aligned}\end{equation}
Here $\Rey \equiv \lambda L U/\nu$ is the Reynolds number, $\Cn \equiv{{\ell_c}}/{L}$ is the Cahn number, and ${\ell_c} =
\sqrt{\nu/\alpha}$ is the length scale above which fluctuations in the homogeneous disordered state ${\bm u}=0$ are linearly
unstable.

\section{\label{sec:vortex} Vortex Solution}
\begin{figure}[!h]
    \centering
    \includegraphics[width=\linewidth]{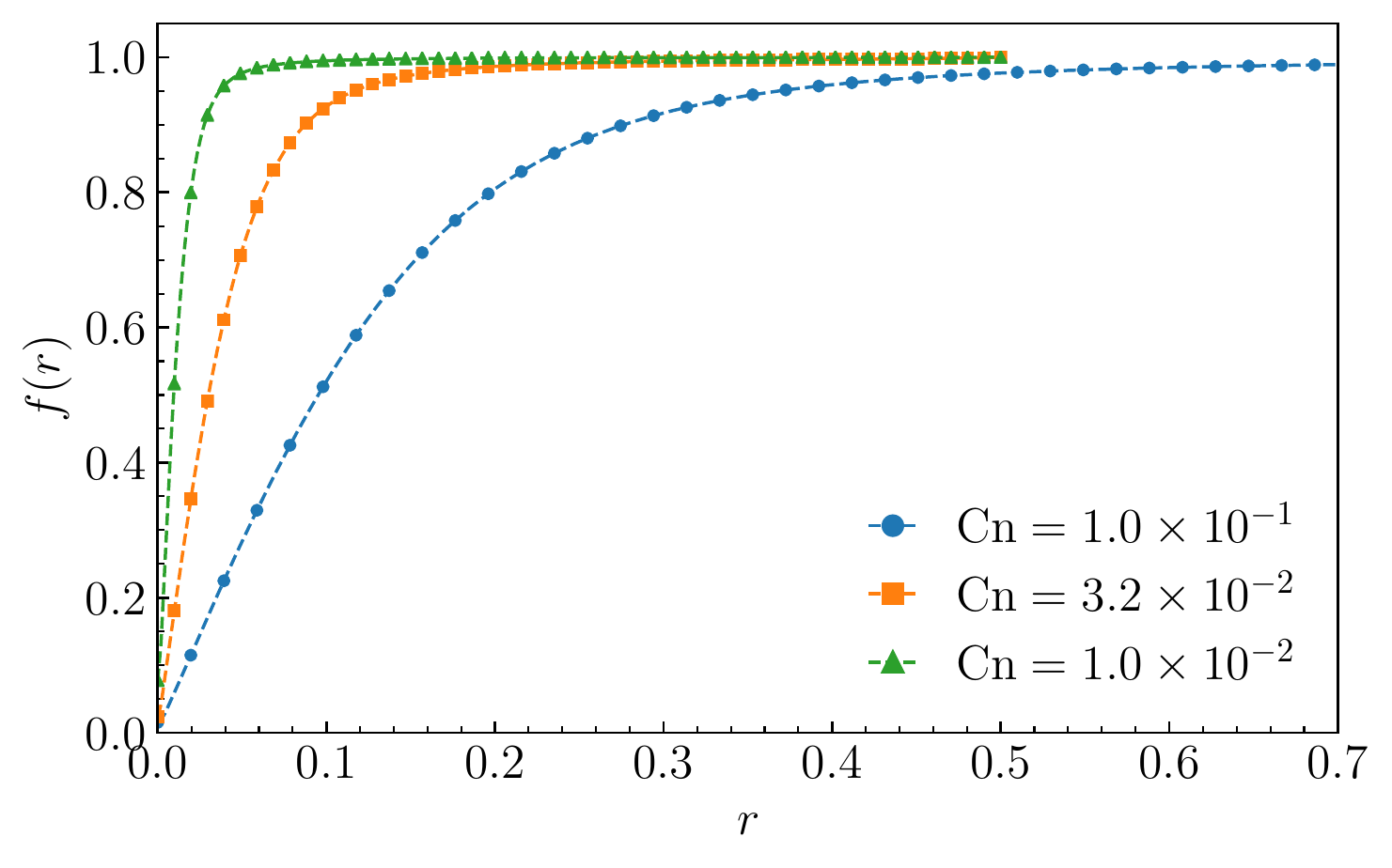}
    \caption{ \label{fig:fr_various_nu}
        Plot of $f(r)$ vs $r$ for different values of $\Cn$.
    }
\end{figure}

Consider the radially symmetric velocity field of an isolated unbounded vortex ${\bm u}({\bm x},t)\equiv f(r) \hat{\theta}$,
where $\hat{\theta}$ is the unit vector along the angular direction, $f(0)=0$, and $f^\prime(1)=0$. Substituting in the ITT
equation, we get the following equations:
\begin{eqnarray}
    \label{eq:s1}
    \left(f^{\prime \prime} + \frac{f^\prime}{r} - \frac{f}{r^2} \right) &=& \frac{1}{\Cn^2}(f^2-1)f , \\
    \label{eq:s2}
    P &=& \Rey\Cn^2 \int_0^r \frac{f^2(r^\prime)}{r^\prime} dr^\prime,
\end{eqnarray}
where the prime indicates the derivative with respect to $r$. Note that \eqref{eq:s1} does not depend on $\Rey$ and is identical
to the equation of a defect in the Ginzburg-Landau equation \cite{Onu02}. In \cref{fig:fr_various_nu} we plot the numerical
solution of $f(r)$ for different values of $\Cn$. For $\Cn<<1$, a regular perturbation analysis reveals that $f(r)\to A r
(1-r^2/8\Cn^2)$.

\begin{figure*}[!t]
    \centering
    \includegraphics[width=0.9\textwidth]{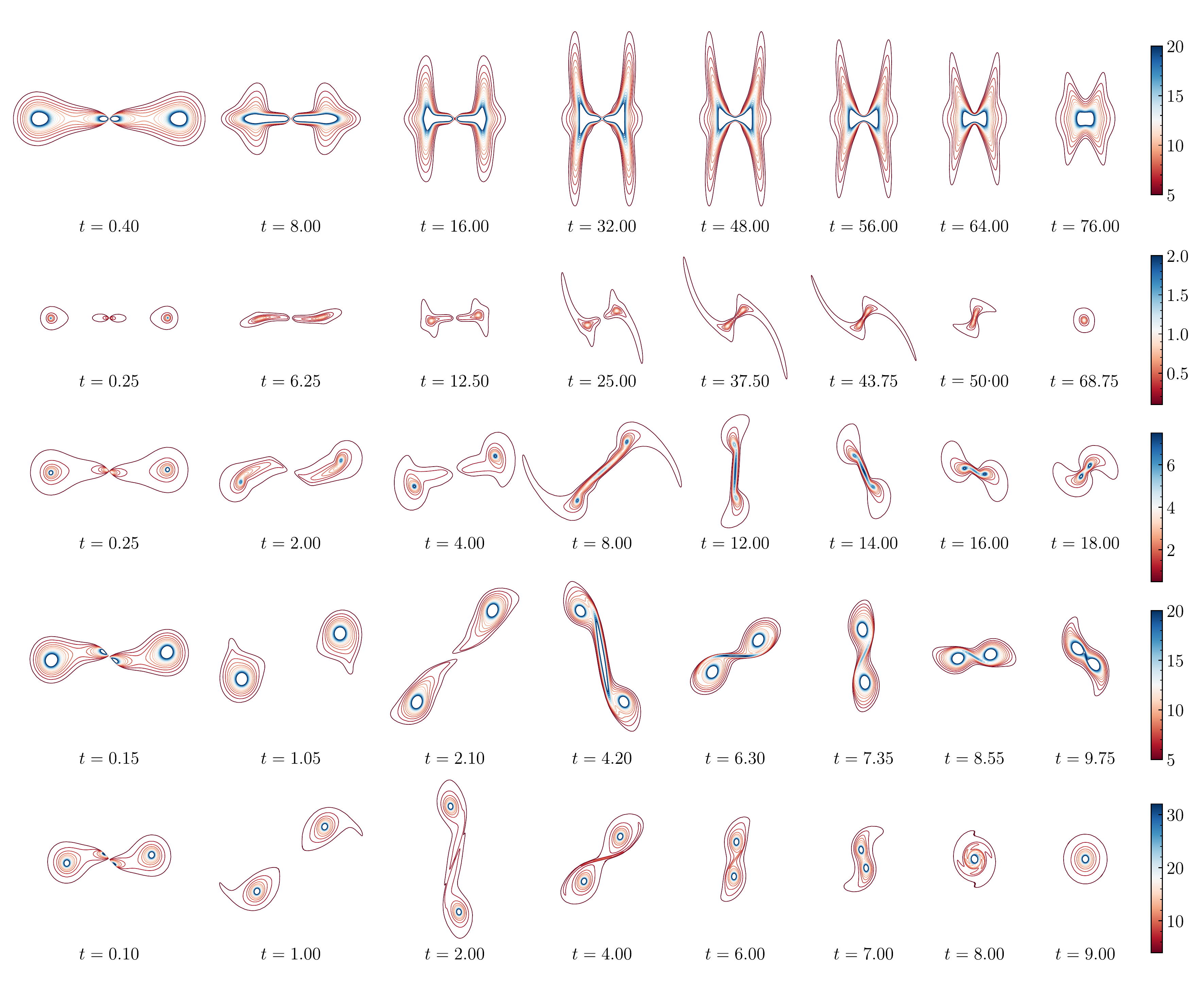}
    \put(-480,365){\Large (a)}
    \put(-480,280){\Large (b)}
    \put(-480,215){\Large (c)}
    \put(-480,135){\Large (d)}
    \put(-480,050){\Large (e)}

    \caption{\label{fig:contours}
        (a)-(e) Contour plots of the vorticity field $\omega$ at various times during the merger process for different values of
        the Reynolds number $\Rey = 0,\, 2\pi \times 10^2,\, 2\pi \times 10^3,\, \pi \times 10^4,\, \rm{and}\; 2\pi \times
        10^4$.
    }
\end{figure*}

\section{\label{sec:merger} Vortex Merger Dynamics}

To investigate the merger of two corotating vortices, we perform a DNS of an isolated vortex-saddle-vortex configuration at
various Reynolds numbers. We use a square domain of area $L^2=4\pi^2$ and discretize it with $N^2=4096^2$ collocation points.
Furthermore, to minimize the effect of periodic boundaries, we set $\alpha=-10$ for $r>0.9 L/2$ and keep $\alpha=1$ otherwise,
where $r\equiv \sqrt{(x - L/2)^2 + (y-L/2)^2}$. This ensures that the velocity decays to zero for $r\geq0.9L/2$. The initial
condition constitutes a saddle at the center of the square domain and two vortices placed at coordinates $[(L-1)/2,L/2]$ and
$[(L+1)/2,L/2]$. As discussed in the main text, it is important to note that (i) similar to the GL equation
\cite{yurke93,Onu02}, vortices in ITT have a topological charge and (ii) similar to the NS equation \cite{Doe95}, the ITT
equation has an advective nonlinearity and the presence of pressure leads to nonlocal interactions.

In \cref{fig:contours}(a)-(e), we plot vorticity contours during different stages of the vortex merger for different $\Rey$.
Since the saddle is at an equal distance away from the two vortices, its position does not change during evolution. For low
$\Rey=0$, the vortex dynamics has similarities to the overdamped motion of defects with opposite topological charge in the
Ginzburg-Landau equation. Vortices get attracted to the saddle and move along a straight-line path. On increasing $\Rey \geq
2\pi \times 10^2$, similar to Navier-Stokes, advective nonlinearity in the ITT becomes crucial. Not only are the vortices
attracted to the saddle, but they also go around each other.

\begin{figure*}[!t]
    \centering
    \includegraphics[width=\linewidth]{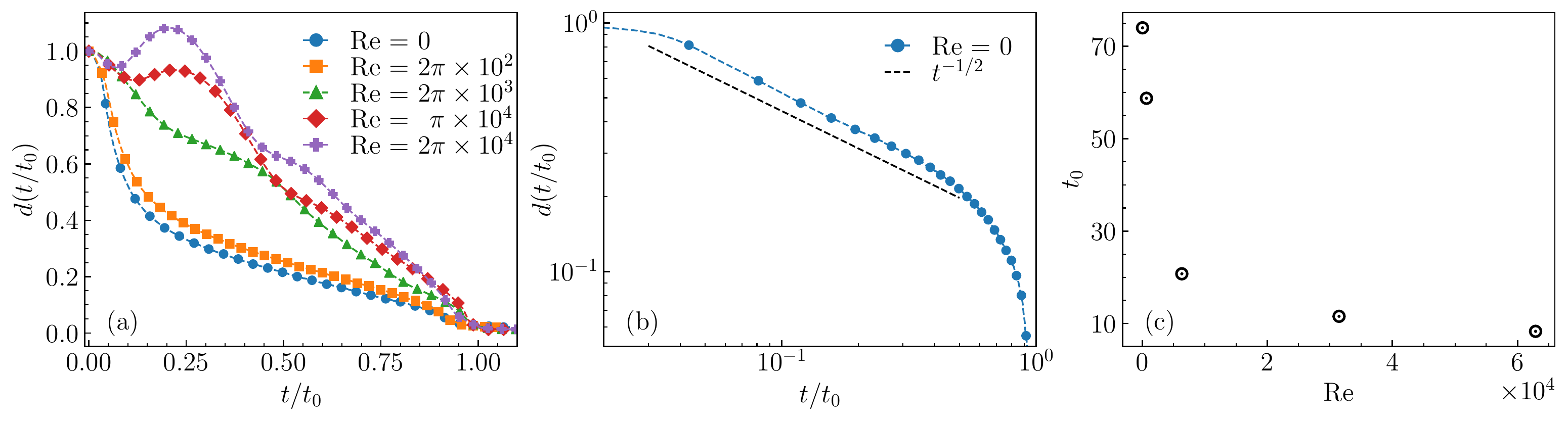}
    \caption{\label{fig:distance}
        (a) Plot of inter vortex distance $d(t)$ vs time $t$ at various Reynolds numbers. The time axis is scaled by the merger
        time $t_{0}$. (b) Log-log plot of $d(t)$ vs $t$ for $\Rey=0$; the black dashed line shows the $1/\sqrt{t}$ scaling. (c)
        Plot of merger time $t_{0}$ versus $\Rey$. As $\Rey$ increases, merger time decreases.
    }
\end{figure*}

In \cref{fig:distance}(a) we plot the intervortex separation $d(t)$ versus time for different $\Rey$. Because of long-range
hydrodynamic interactions due to incompressibility, the merger dynamics is accelerated even for $\Rey=0$. The intervortex
separation decreases as $d(t)\sim 1/\sqrt{t}$ [see \cref{fig:distance}(b)], in contrast to the much slower $d(t)\sim
\sqrt{t_0-t}$ observed in the GL dynamics \cite{lubensky1995,denniston1996}. On increasing the $\Rey$ number, inertia becomes
dominant, vortices rotate around each other, and $d(t)$ decreases in an oscillatory manner. The time for the merger $t_0$
decreases with increasing $\Rey$ [see \cref{fig:distance}(c)].

\clearpage

 
%
\end{document}